\documentclass[aps,prb,reprint,superscriptaddress, showkeys]{revtex4-2}
\usepackage{graphicx}
\usepackage{dcolumn}
\usepackage{bm}
\UseRawInputEncoding
\usepackage{chemformula} 
\usepackage[T1]{fontenc} 
\usepackage{amsfonts, amsmath, amsthm, amssymb}
\usepackage{hyperref}
\usepackage[mathlines]{lineno}

\usepackage[dvipsnames]{xcolor}

\begin{document}
\title{Integration of \texorpdfstring{$\text{Er}^{3+}$}{Er3+}-emitters in silicon-on-insulator nanodisks metasurface}
\author{Joshua Bader}
\affiliation{School of Engineering, RMIT University, Melbourne, 3000, VIC, Australia.}
\affiliation{Quantum Photonics Laboratory and Centre for Quantum Computation and
Communication Technology, School of Engineering, RMIT University, Melbourne, 3000,
VIC, Australia.}
\author{Hamed Arianfard}
\affiliation{School of Engineering, RMIT University, Melbourne, 3000, VIC, Australia.}
\affiliation{Quantum Photonics Laboratory and Centre for Quantum Computation and
Communication Technology, School of Engineering, RMIT University, Melbourne, 3000,
VIC, Australia.}
\author{Vincenzo Ciavolino}
\affiliation{School of Engineering, RMIT University, Melbourne, 3000, VIC, Australia.}
\author{Shin-ichiro Sato}
\affiliation{Quantum Materials and Applications Research Center (QUARC),
National Institutes for Quantum Science and Technology (QST), Takasaki,  370-1292, Gunma, Japan
}
\author{Stefania Castelletto}
\affiliation{School of Engineering, RMIT University, Melbourne, 3000, VIC, Australia.}
\email{stefania.castelletto@rmit.edu.au}



\begin{abstract}
Erbium ($\text{Er}^{3+}$) emitters are relevant for optical applications due to their narrow emission line directly in the telecom C-band due to the ${}^\text{4}\text{I}_{\text{13/2}}$ $\rightarrow$ ${}^\text{4}\text{I}_{\text{15/2}}$ transition at 1.54 $\mu$m. Additionally they are promising candidates for future quantum technologies when embedded in thin-film silicon-on-insulator (SOI) to achieve fabrication scalability and CMOS compatibility. 
In this paper we integrate $\text{Er}^{3+}$ emitters in SOI metasurfaces made of closely spaced array of nanodisks, to study their spontaneous emission via room and cryogenic temperature confocal microscopy, off-resonance and in-resonance photoluminescence excitation at room temperature and time resolved spectroscopy. 
This work demonstrates the possibility to adopt CMOS-compatible and fabrication scalable metasurfaces for controlling and improving the collection efficiency of the spontaneous emission from the $\text{Er}^{3+}$ transition in SOI and could be adopted in similar technologically advanced materials.  
\end{abstract}
\maketitle
\section{Introduction}

Rare-earth ions in various crystals have been recognized in spectroscopy since mid-nineteenth century, and from the material science observation domain, it is now possible to deploy their properties in optical technologies, such as laser, displays, fiber optical communication \cite{miniscalco1991erbium,digonnet2001rare,kenyon2002recent}, integrated photonics \cite{liu2021chip,LiuphoEramp2022} and more recently in quantum technologies\cite{stevenson2022erbium, Becher_2023}.
The key appeal of rare-earth ions, from single ions \cite{yin2013optical,Songtaosingleshot2020} right through to large ensembles \cite{Ahlefeldtultranarrow2026}, is the high optical and spin coherence with narrow 4f-4f optical transition homogeneous line-widths at liquid He temperatures, as narrow as 73 Hz or coherence lifetime T$_2$ >4 ms, in $\text{Er}^{3+}$:Y$_2$SiO$_5$ single crystals\cite{PhysRevB.73.075101,rakonjac2020long}, $\text{Er}^{3+}$ in Y$_2$O$_3$ \cite{Songtaosingleshot2020,gupta2023robust}, in Si \cite{berkman2025long, yin2013optical} and implanted in Si waveguides and photonic cavity\cite{gritsch2022narrow,Gritsch:23}. 

In particular $\text{Er}^{3+}$ in silicon-on-insulator (SOI) is a promising material platform as it is a complementary metal-oxide-semiconductor (CMOS) compatible and leverages well-established nanofabrication methods to enable scalable integration of quantum networks and quantum memories \cite{Bhaskar2020}. 
One of the key limitations of $\text{Er}^{3+}$ in Si and SOI is its very low emission rate, arising from the long optical lifetime (on the order of $\mu$s or ms), low quantum efficiency (a few percent), and the high refractive index of the material, which limits photon extraction. Therefore surface nanostructuring is required to enhance the collection efficiency of the $\text{Er}^{3+}$ optical transition at 1534 nm. 

The use of all high-index dielectric Mie-resonant metamaterials based on Si has been suggested as a viable platform for enhancing the fluorescent emission\cite{Babicheva:24}. $\text{Eu}^{3+}$, another trivalent lanthanide ion emitting in the visible range, has shown a maximum absolute averaged fluorescence emission enhancement by a factor of 6.5 by using dielectric metasurfaces composed of Mie-resonant silicon nanocylinders\cite{Vaskinmaniupulation2019}. 

In this work, we focus on using facile fabrication approaches to achieve scalability of the metasurfaces, by directly fabricate them in SOI with the $\text{Er}^{3+}$ derived by ions implantation. Further here we investigate the photoluminescence excitation of the $\text{Er}^{3+}$ in SOI at room temperature.

We fabricated the SOI nanodisks array metasurfaces with implanted $\text{Er}^{3+}$ at the centre of thin film, to study their ensemble fluorescence enhancement. 
We report the successful integration of ensemble of $\text{Er}^{3+}$-defects in SOI metasurfaces by means of studying the $\text{Er}^{3+}$ photoluminescence (PL) at room temperature (RT) and cryogenic temperatures, the PL time-resolved lifetime measurements, the absorption and emission PL polarization properties and the room temperature photoluminescence excitation (PLE). 

By using confocal microscopy and time-resolved PL at RT, we observed an average maximum experimental PL enhancement of a factor of 5 around the 1535$\pm$3 nm transition. The previous implantation of Er in SOI nanopillars provided a fluorescence enhancement of a factor of 2 \cite{Takahashi2023}, while a factor of 1.4 enhancement was achieved in Erbium infused silicon nanocones on silica \cite{yaroshenko2023active}, providing an improvement using SOI metasurfaces.

We measured the PL optical lifetime of the $\text{Er}^{3+}$ transition in the metasurfaces and in the unpatterned SOI, with a maximum lifetime reduction of 9.6\%. By PLE we observe a 2.9 enhancement and a factor of 2 lifetime reduction for a transition at 1534.79 nm. 
This work paves the way for the utilization of scalable SOI metasurfaces to achieve $\text{Er}^{3+}$ emission enhancement through fine-tuning of the metasurfaces' design.

\section{Materials and Methods}
\subsection{\texorpdfstring{$\text{Er}^{3+}$}{Er3+} implantation}\label{sec8}
The ions implantation was performed using a 400 kV ion implanter at the Takasaki Institute for Advanced Quantum Science, QST. The implantation of 350 keV Er ions at a fluence of $4.0\times10^{12}\,\text{cm}^{-2}$, followed by the implantation of 50 keV O ions at a fluence of $3.0\times10^{13}\,\text{cm}^{-2}$ was performed at a normal angle at room temperature. The ion beams were raster scanned to uniformly implant these ions into the sample. We simulated the Er- and O-ion concentration depth profiles with the stopping and range of ions in matter (SRIM) software package \cite{ZIEGLER20101818}, and the implantation energies were chosen to achieve maximal overlap with the metasurfaces. The implanted Er ions were not isotopically separated. 
\subsection{Dielectric Metasurface fabrication} \label{sec2}
\begin{figure*}
\includegraphics[width=\textwidth]{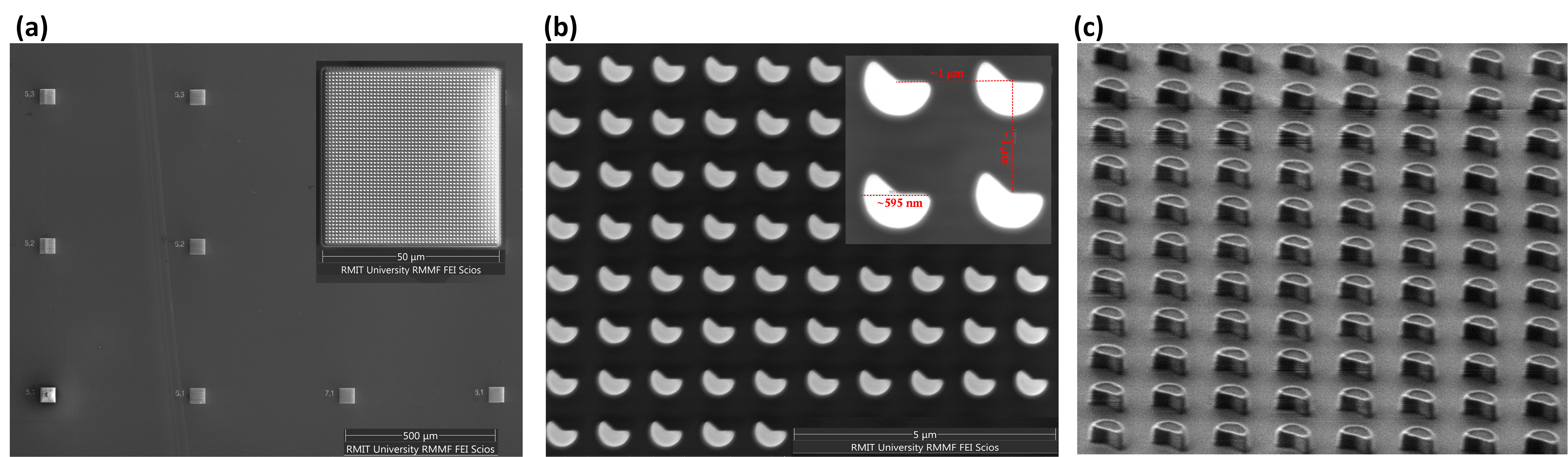}
\caption{(a) Top-view SEM image of the fabricated device showing an array of 50 $\times$ 50 $\mu$m$^2$ square metasurface elements. The inset provides a zoomed-in view of one of the square metasurface elements, revealing its detailed structure. (b) Higher magnification SEM view of (a), highlighting the symmetry-breaking nanodisk array forming the metasurface. The inset further illustrates the shape and dimensions of the individual symmetry-breaking nanodisks. (c) Cross-sectional FIB image of the fabricated nanodisk array, captured at a 52$^\circ$ tilt to reveal the height and structural arrangement of the asymmetric nanodisks.} \label{fig3} 
\end{figure*}

The metasurfaces were fabricated on an implanted SOI-sample with a 260 nm thick top silicon layer (type P/B, Orientation: <100>, resistivity 14-22 ohm $\cdot$ cm)
 and a 2 $\mu$m buried oxide (BOX) layer with nanodisks diameter of 595 nm, pitch of 1 $\mu$m and sector cut-angle of 138$\pm$2$^{\circ}$, based on a previous design methodology predicting theoretical Mie resonant scattering modes collective behavior \cite{AhamadSiC2024,AHAMAD2025120881}.
The fabrication process largely adheres to standard CMOS techniques, except for the patterning step, which was carried out using electron-beam lithography (EBL). Specifically, a Vistec EBPG 5200 electron-beam lithography system was used to define the device pattern on a positive photoresist (ZEP520A). The patterned resist was then transferred to the top silicon layer through inductively coupled plasma (ICP) reactive ion etching (RIE), utilizing SF$_6$ and C$_4$F$_8$ as the etching gases. Figure \ref{fig3}(a) shows a top-view scanning electron microscopy (SEM) image of the fabricated device, which consists of an array of square metasurface elements, each measuring 50 $\times$ 50 $\mu$m$^2$. A higher magnification SEM image of a single metasurface element (Figure \ref{fig3}(b)) reveals the periodic nanodisks array forming the metasurface. The inset further highlights the precise geometry of the nanodisks, confirming their well-defined shape and uniformity. Figure \ref{fig3}(c) presents a cross-sectional focused ion beam (FIB) image of the fabricated nanodisks array, captured at a 52$^\circ$ tilt to reveal the vertical profile and spatial arrangement of the nanodisks. 
\subsection{Metasurfaces annealing}\label{sec12} 
In order to optically activate the Er-O defect, thermal annealing was carried out after metasurfaces' fabrication at 500\textdegree C \cite{rinner2023erbium} and 700\textdegree C  \cite{yin2013optical} for 1 minutes and 10 minute in $\text{N}_{2}$ atmosphere, respectively, to study the effect of annealing. 
\subsection{Photoluminescence spectroscopy}\label{sec14}
The spectroscopy investigation was carried out using a custom built confocal microscope with a Thorlabs 976-P300 continuous-wave laser-diode for excitation, separating the PL from the laser by a 980 nm longpass dichroic mirror. A Olympus LC Plan 0.65 NA 50$\times$ for the low-temperature investigation) or a 0.85 NA 100$\times$ dry objective (for RT investigation) were used to direct the excitation laser onto the investigated samples and collect back the fluorescence. Higher NA and higher magnification provided the highest enhancement recorded. Furthermore, a full polarizer (FP) and a $\lambda$/2 waveplate (HWP), at 976 nm, were present on the excitation  during this study and set to achieve the maximum PL.
A Montana cryostation operating with a closed-loop helium cycle was employed to study the low temperature properties of the Er-O defect. An ID Quantique InGaAs avalanche-photodiode (APD) as well as a Princeton Instruments $\text{LN}_{2}$-cooled spectrometer were utilized to detect the fluorescence infrared photons. Details of the experimental setup are provided in Figure S1 of the \textit{Supporting Information}.
\subsection{Confocal imaging}\label{sec13}
We used another custom built confocal microscopes to acquire the high resolution metasurfaces' images equipped with a 780 nm continuous-wave laser, a 900 nm long pass dichroic mirror (DM) followed by another 900 nm long pass (LP) filter and an objective with 100$\times$ magnification and 0.85 NA.
The fluorescence is focused by an achromatic doublet lens into a single mode (SM) fiber that acts as a pinhole. The detector used is a Single Quantum EOS-810 superconducting nanowire single-photon detector (SNSPD). The sample was positioned on a PI NanoCube piezo stage with 100$\times$100 $\mu$m travel range. Details of the experimental setup are provided in Figure S2 of the \textit{Supporting Information}.
\subsection{Time-resolved photoluminescence measurements}
A Thorlabs MC1F2 optical beam-chopper driven by a arbitrary waveform generator was implemented into the excitation section of the confocal microscope (Figure S1 of the \textit{Supporting Information}), which modulated the applied 976 nm excitation for the optical lifetime investigation. A 1550 $\pm$ 50 nm bandpass (BP) isolated the $\text{Er}^{3+}$-emission.
\subsection{Polarization Measurements}
One FP and one HWP were placed into the emission section of the experimental setup (Figure S1 of the \textit{Supporting Information}), tailored to the respective wavelength. The already present FP, HWP and BP remain within the excitation/emission- section of the experimental setup respectively to investigate the polarization of the observed defect.
\subsection{Photo-luminescence excitation and time resolved in resonance excitation.}
PLE spectroscopy is carried out in a free-space configuration at RT. Tunable excitation (Cobrite DX100), ranging from 192.7924 THz to 195.9439 THz (1530 nm and 1555 nm),  is modulated (acoustic-optic modulator) and then applied via an Olympus 100X Objective (LCPlan N 100x 0.85 NA) in 500 MHz increments with photon events captured subsequently by an SNSPD. During excitation, the bias current from the SNSPD is set to zero and established back to a valid value at a time-range which coincidences with the falling edge of the excitation. The falling edge of the excitation is simultaneously detected and timestamped (ID Quantique ID801) which is seen as a trigger event.
Following that, the incoming timestamps of the photon events are recorded. Via post-processing, we remove all timestamped photon-events before the falling edge trigger and are able to extract the amount of detected photons for each applied pulse 1 ms after the trigger event. 
Resonant optical lifetime transients are obtained by first identifying the inhomogeneous resonances within the PLE-spectrum via single Gaussian fitments. Following that, the excitation is configured to drive the determined peak with a subsequent transient acquisition for the required time. Finally, a bi-exponential fit reveals the optical lifetime.
\section{Results and Discussions}
\subsection{Photoluminescence modification}\label{sec4}
\begin{figure*}[htbp!]
\includegraphics[width=\textwidth]{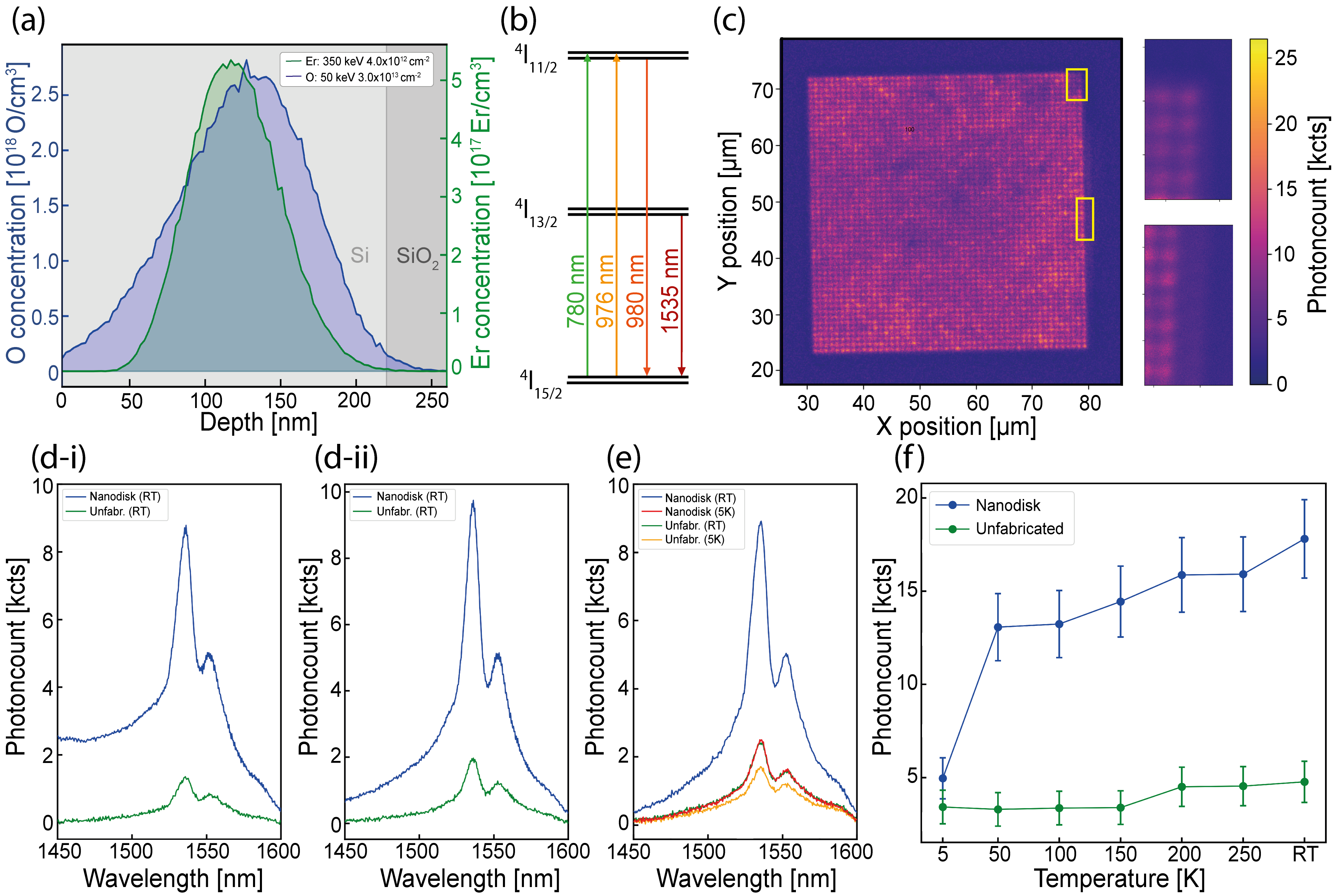}
\caption{$\text{Er}^{3+}$ ion implantation and photoluminescence properties: (a) Er (green) and O (blue) implantation profile in a 1:10 ratio, focusing on defect generation within the Si-layer. (b) Energy level diagram illustrating all excited transitions \cite{larin2021luminescent}. (c) Obtained confocal-microscope scan illustrating the fabricated nanodisks array with 1 mW excitation at 780 nm in combination with a 900 nm longpass filter (LP). (d) Off-resonance PL-spectrum obtained for defects within the metasurface or within unfabricated (unfabr.) sections using a high NA 100X-objective, and 2mW - 976 nm excitation considering a 500 \textdegree C anneal in (i) and  700 \textdegree C anneal in (ii). 
(e) Off-resonance PL-spectrum obtained using a lower NA objective for defects within the metasurface or within unfabricated sections at 5K and RT, respectively with 2mW - 976 nm excitation. (f) $\text{Er}^{3+}$ transition line traces observed over different measurement temperatures. Data points are determined by integrating the individually obtained $\text{Er}^{3+}$ transition line spectra at 1535 nm over 3 nm around the central peak. The illustrated error-bars were defined as the square-root of the determined integral.} 
\label{fig4} 
\end{figure*}

We investigated the SOI-metasurfaces with Er-O defects introduced via ion implantation in a 1:10 (Er:O) ratio\cite{benton1991electrical, coffa1993optical,lourencco2016super} during a two-step implantation process: (1) 350 keV $4.0 \times 10^{12}$ $\text{Er}/\text{cm}^{2}$ and (2) 50 keV $3.0 \times 10^{13}$ $\text{O}/\text{cm}^{2}$ as shown in Figure~\ref{fig4}(a). The relevant energy level scheme is illustrated in Figure~\ref{fig4}(b), showing the Er$^{3+}$ ${}^\text{4}\text{I}_{\text{13/2}}$ $\rightarrow$ ${}^\text{4}\text{I}_{\text{15/2}}$ transition. \textcolor{black}{However, due to off-resonance excitation, we also excite the ${}^\text{4}\text{I}_{\text{11/2}}$ $\rightarrow$ ${}^\text{4}\text{I}_{\text{15/2}}$ transition.}

The samples were imaged utilizing custom confocal microscopes (see Methods) using 976 nm excitation before and after annealing, and using 780 nm excitation after annealing (as shown in Figure~\ref{fig4}(c)). Details of the experimental setups are provided in the \textit{Supporting Information}. The confocal images revealed bright areas where the nanodisks were fabricated. 

We observed a photon-emission increase after the performed annealing at temperatures of 500 \textdegree C  \cite{rinner2023erbium} and 700 \textdegree C \cite{yin2013optical}, compared to the not annealed samples. This is in agreement with previous studies showing that the Er-defects cause the O-ions to redistribute within the implanted layer, leading to Er-O defects via ionic bonding \cite{favennec1990optical}, which can evolve into electron traps \cite{kenyon2005erbium}. 

We investigated the spectroscopy properties of the  Er$^{3+}$ ${}^\text{4}\text{I}_{\text{13/2}}$ $\rightarrow$ ${}^\text{4}\text{I}_{\text{15/2}}$ transition at 5 K and at RT, for both, emitters within the nanodisks array and in the unfabricated sections of the samples. 
Our findings illustrated in Figs.~\ref{fig4}(d-i), (d-ii) show that the ${}^\text{4}\text{I}_{\text{13/2}}$ $\rightarrow$ ${}^\text{4}\text{I}_{\text{15/2}}$ transition line is present
in both considered annealing steps,  utilizing a high NA objective at RT. A measured PL-enhancement with a factor of 5 can be determined, considering only the $\text{Er}^{3+}$-transition line at 1535$\pm$3 nm.

By performing lower temperature spectroscopy, for both cases with a lower NA objective, we observed a strong PL-quenching at temperatures below 50 K, while the PL tends to be more stable at higher temperatures, as illustrated in Figure~\ref{fig4}(e). 

Regardless, we still recorded a PL enhancement of 3.7 at cryogenic temperature, considering only the  $\text{Er}^{3+}$-transition line at 1535 nm. We also observed a weaker shoulder of the emission between 1550 nm and 1552 nm, which can be attributed to Stark splitting effects \cite{miritello2006optical, li2012photoluminescence}. 
Further comparisons at other temperatures led to a enhancement factor reduced to 1.45 at 5 K, as illustrated in Figure~\ref{fig4}(f).
The lower enhancement observed at 5 K could be justified by the here high oxygen co-doping, inducing different Er-O centres emission as observed in ref. \cite{Hong:21} with similar Er:O ratio. Our work is not focusing on the study of the PL temperature effects rather PL enhancement effects, and we observed higher enhancement at RT with associated reduced thermal quenching. 
By applying a larger integration bandwidth, an enhancement reduction is observed as shown in Figure S3 of the \textit{Supporting Information}.

To better resolve the internal structure of the fabricated metasurfaces, we used a confocal fluorescence microscope equipped by a SNSPD, as described in the Methods (details in \textit{Supporting Information}). Exciting with 1 mW off-resonance at 780 nm and using 900 nm long pass filters for the collection of the emitted photons, we observed an enhanced emission in the fabricated area with respect to the background which is attributed to the ${}^\text{4}\text{I}_{\text{11/2}}$ $\rightarrow$ ${}^\text{4}\text{I}_{\text{15/2}}$ transition line centered at 980 nm. A scan of 60 $\mu m$ $\times$ 60 $\mu m$ has been acquired with a resolution of 500 nm, making it possible to see the array of nanodisks (inset of Figure~\ref{fig4}(c)) and that the emission is actually confined within them. From higher resolution imaging we observed different brightness of the Er emission in the metasurfaces, possibly due to an non uniform activation of the emitters in the whole structure.

\subsection{Time resolved photoluminescence}
\begin{figure*}
\includegraphics[keepaspectratio, width=\textwidth]{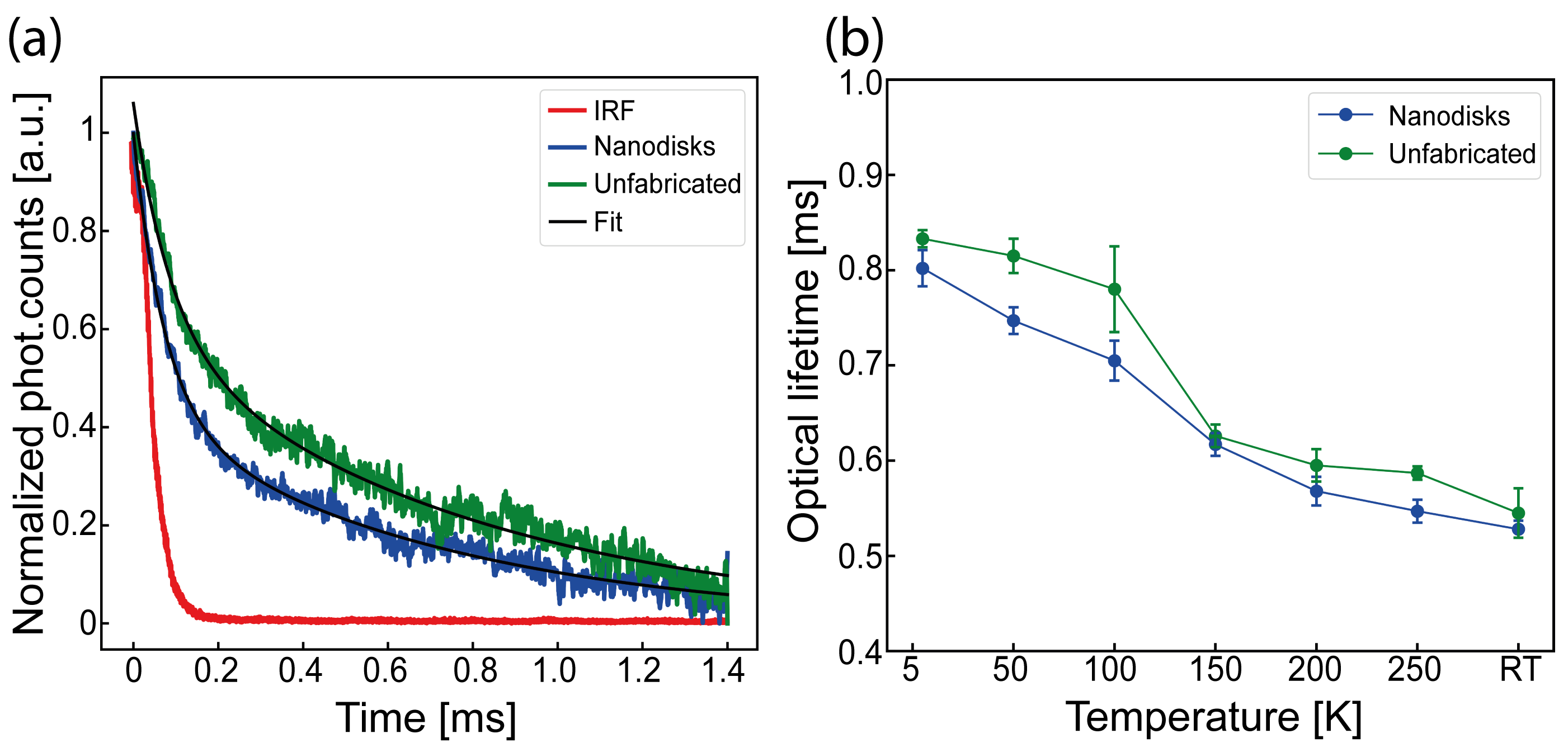}
    \centering
    \caption{\textbf{}Optical lifetime properties of observed the $\text{Er}^{3+}$-transition: (a) observed transients from the Er-O defect at 100 K; (b) Observed optical lifetimes over various measurement temperatures with 976 nm excitation-wavelength in combination with a 1550 $\pm$ 50 nm Bandpass (BP).} 
\label{fig6}
\end{figure*}

We measured the PL time transients from the Er-O defects to determine the optical lifetime in the metasurface and in the unpatterned SOI area. Er related defects are usually known for lifetimes greater than 1 ms even when integrated into photonic devices \cite{weiss2021erbium}. For this investigation, we modulated the applied optical excitation, synchronized the modulation-trigger to the detector output and isolated the relevant emission with a 1550 $\pm$ 50 nm bandpass filter (See Methods and \textit{Supporting Information}). Here, the instrument response function (IRF) as well as the emission transients obtained from Erbium emitters within the nanodisks array and within unfabricated sample-sections, were fitted with a bi-exponential function as following: 

\begin{equation}
    \tau_{\text{Fit}} = \text{a} \cdot e^{-\frac{\text{b}}{\tau_1}} + \text{c} \cdot e^{-\frac{\text{d}}{\tau_2}}
\end{equation}

\noindent utilized for the $\text{Er}^{3+}$ related emission, as shown in Figure~\ref{fig6}(a). We estimated an optical lifetimes of $\tau_{\text{Br.- Sym.}} \approx 528 \pm 9.3$ $\mu s$ as well as $\tau_{\text{Unfabr.}} \approx 545 \pm 26.5$ $\mu s$ at RT.  

We investigated the temperature dependence of the optical lifetime as shown in Figure~\ref{fig6}(b) by conducting several measurements over different temperatures ranging from 5 K to RT with the maximum lifetime observed at lower temperature (5 K), for both scenarios of $\tau_{\text{Br.- Sym.}} \approx 802 \pm 18.5$ $\mu s$ and $\tau_{\text{Unfabr.}} \approx 833 \pm 8.7$ $\mu s$ respectively. 
We observed an overall lifetime-variation of 274 $\mu s$ for emitters embedded in the the nanodisks metasurfaces and 288 $\mu s$ for emitters embedded in unfabricated sections of the sample across all temperatures. The reduction of the lifetime at higher temperature follows the increased PL emission at the same temperature as previously observed with high Oxygen co-doping\cite{Hong:21}.

The largest impact of the nanodisks design on the lifetime can be observed at 100 K with a lifetime reduction of approximately 9.6 \% with specific lifetimes of $\tau_{\text{Br.- Sym.}} \approx 705 \pm 21$ $\mu s$ and $\tau_{\text{Unfabr.}} \approx 780 \pm 44$ $\mu s$ respectively. The lower PL lifetime (<1-2 ms) here observed for the fabricated sample, compared to the unpatterned SOI, can be attributed to a small variation of the local density of electromagnetic states in the nanodisks array due to surface reflectance. 
The spectral reflectance measurement of the metasurface does not in fact show any resonance at 1535 nm based on Mie Scattering modes coherent superposition (see Figure S4 in the \textit{Supplementary information)}, as expected by such nanostructure design based on ref.\cite{AhamadSiC2024} and no evidence of narrow Fano resonances due to the broken symmetry nanodisks is observed, as conversely reported in ref. \cite{campione2016broken, CuiMultiple2018}. This justifies the here low decay rate enhancement of the metasurface, as similarly observed in close-spaced pillars array in silicon carbide \cite{Vuillermet2025}. Nevertheless the close spacing of the nanodisks further increases the collection efficiency, possibly due to Mie scattering modes of individual nanodisks, resulting in an overall stronger reflectance of the surface and enhanced light extraction from the nanodisks \cite{Bezares:13}. It is also to be considered that such Mie scattering nanostructures \cite{AhamadSiC2024, AHAMAD2025120881} or metasurfaces emission rate enhancement strongly depend on the ensemble emitters' dipole orientation, which is difficult to simulate, and from the ensemble of emitters rather than single emitters. As such careful design should be performed to achieve Purcell's enhancement, as this has not been demonstrated as yet in broken-symmetry Mie scattering metasurfaces with embedded quantum dots \cite{liu2018light,CuiMultiple2018}. Finally it has to be noted that the simulation results presented in ref. \cite{AHAMAD2025, AHAMAD2025120881} are based on idealized material parameters and nanostructure geometries, while sensitivity studies  to deviations in structural dimensions  may explain  differences between simulated and measured performance, with reduced Q-factors and resonance shifts in experimental implementations.

\subsection{Metasurface absorption and emission dipole polarization }\label{sec5}
\begin{figure}
\centering
\includegraphics[keepaspectratio, width=9cm]{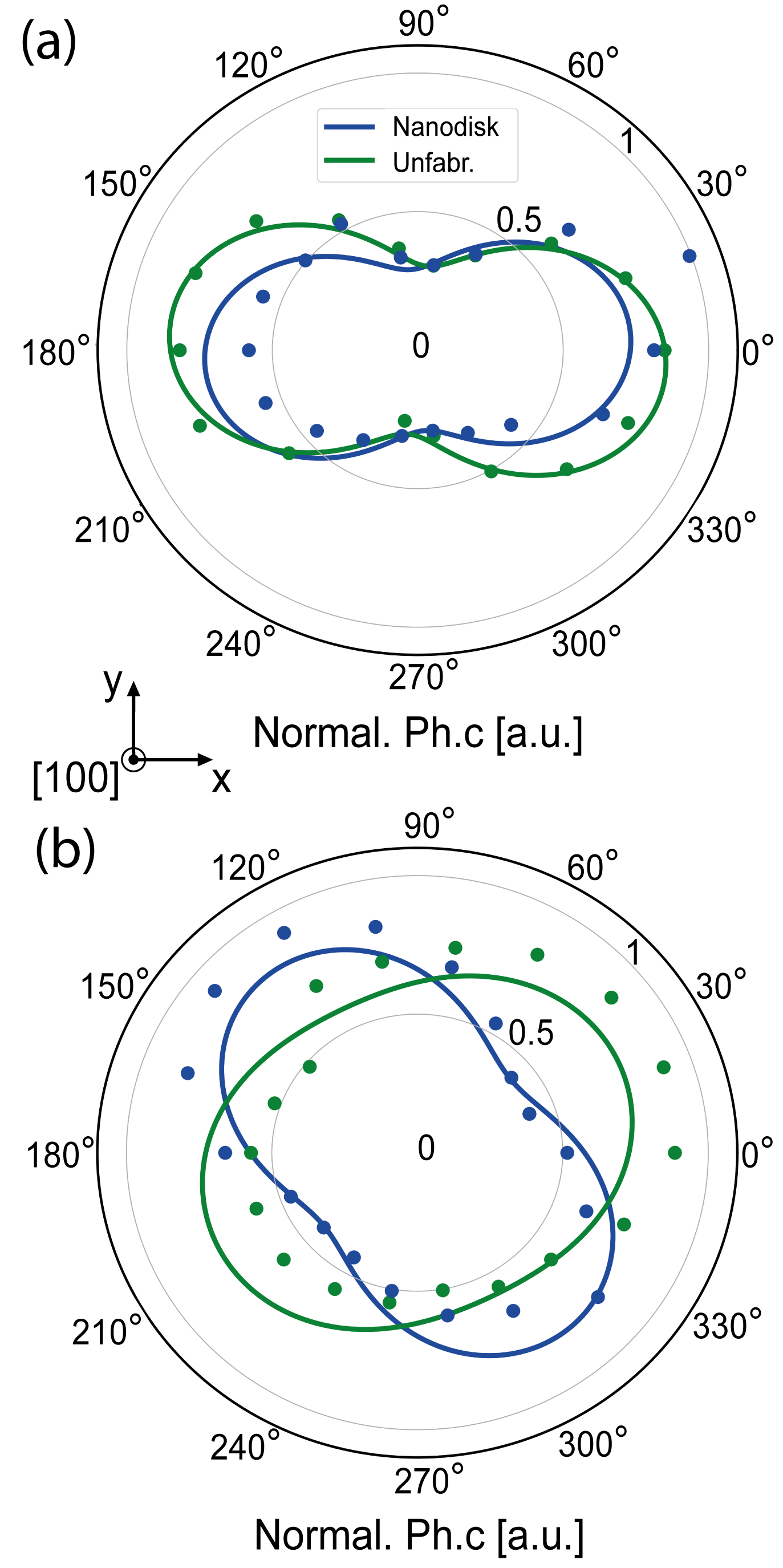}    
\caption{Polarization properties of observed $\text{Er}^{3+}$-defects: (a) observed photon absorption dipole. (b) observed photon emission properties from the ensemble-defect.}
\label{fig8}
\end{figure}

As illustrated in Figure~\ref{fig8}(a), the nanodisks metasurface has only minor impact on the absorption dipole behavior from the Er-O defects determined by rotating a 976 nm half-wave plate (HWP) in combination with an full polariser (FP) win the excitation beam, as shown in the experimental setup of Figure S1 in the \textit{Supporting Information}. We perform a fitting of the measured data based on
\begin{equation}
     I = a + b \cdot sin^2(\theta_{\text{HWP}}+\phi),
\end{equation}

with $a, b$ and $\phi$ as fit parameters \cite{wang2020experimental}.
Polarization visibilities of $\eta_{\text{Abs., Br.- Sym.}}$ $\approx$ 53.2 \% and $\eta_{\text{Abs., Unfabr.}}$ $\approx$ 54.2 \% respectively can be identified, calculated as \cite{wang2020experimental}
\begin{equation}
    \eta = \frac{I_{\text{max}} - I_{\text{min}}}{I_{\text{max}} + I_{\text{min}}}.
\end{equation}
Furthermore, we identify unpolarized dipole behavior when studying the emission properties of Er-O defects within unfabricated sections of the sample by rotating a 1550 nm HWP in combination with a FP while the excitation section is set to its respective maximum, as shown in Figure~\ref{fig8}(b). During this study, a 1550 $\pm$ 50 nm BP isolated the relevant emission. A polarization visibility $\eta_{\text{Em., Unfabr.}}$ of 30.2 \% can be identified. Interestingly, a polarization visibility $\eta_{\text{Em., Br. - Sym.}}$ of 38.2 \% can be determined in the opposing studied case. We attribute this 8 \%-emission polarization increase to the effects of the nanodisks array. That aligns with the previous presented findings where a photoluminescence enhancement was identified and caused by an increased alignment of the emission dipole with respect to the [100]-axis of the crystal within the presented metasurface. 

\subsection{Room temperature photo-luminescence excitation and resonant time resolved emission}\label{PLE}
\begin{figure*}
\centering
\includegraphics[width=\textwidth]{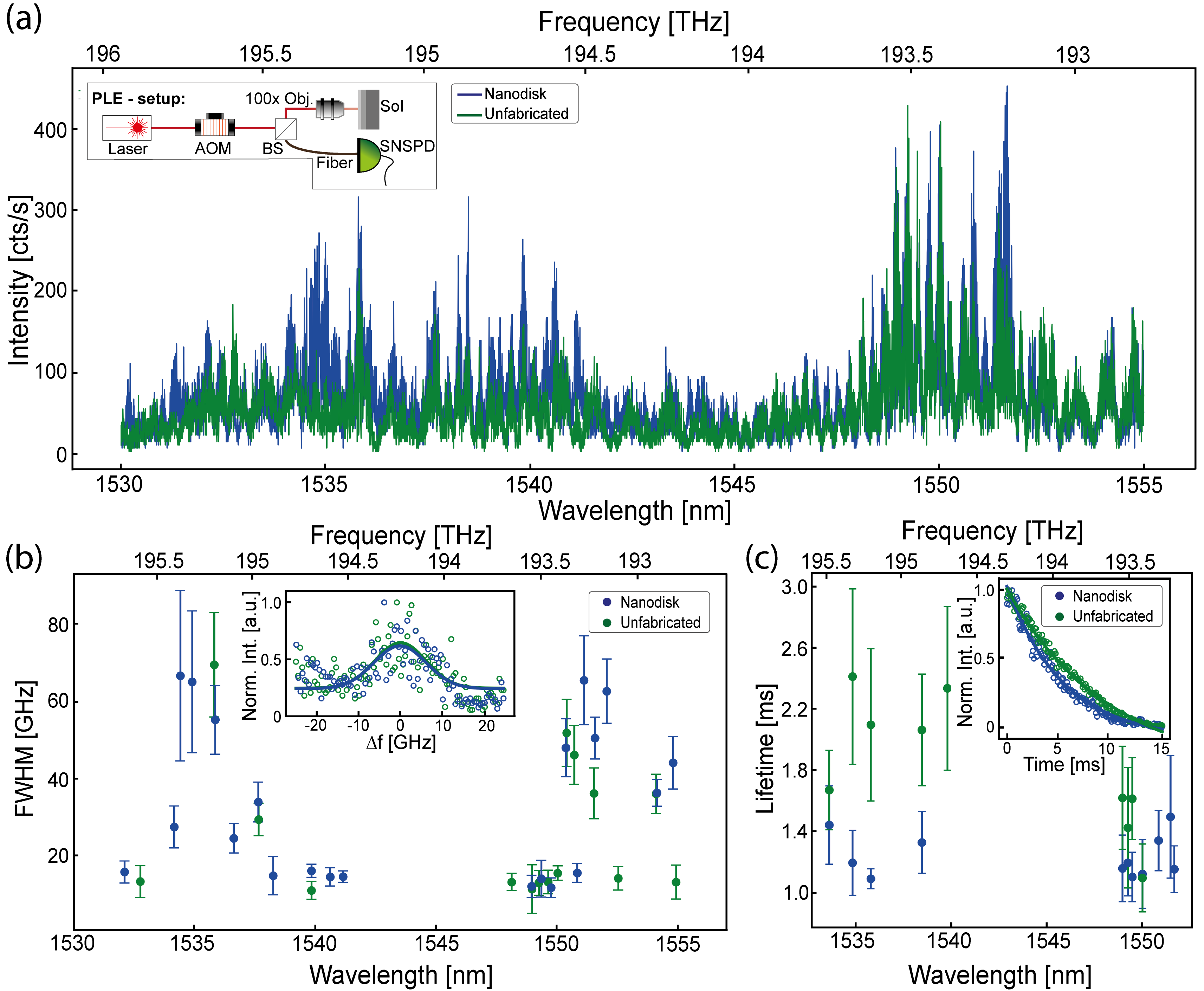}    
\caption{PLE investigation at room temperature: (a) obtained photoluminescence excitation spectrum from the Er-O implanted SOI-samples in both investigated areas (nanodisks and unfabricated). The inset illustrates the utilized experimental setup schematically;  (b) exhibited inhomogeneous linewidths for all detected significant Er-O defect resonances in relation to observed wavelength where dots represent the linewidth with subsequent fitment-uncertainties illustrated as vertical error-bars. The narrowest identified  resonances are shown within the inset for both, the nanodisks and the unfabricated area where a single Gaussian fit (solid lines) is applied onto the measured data (dotted); (c) lifetime-decay overview illustrated with dots in relation to observed resonance-wavelength. Subsequent fitment-uncertainties are illustrated as vertical error-bars. Normalized measured decay-transients obtained over 15 ms (dotted) with subsequently applied bi-exponential fit (solid lines) are shown within the inset from the 1534.84 nm-resonance.
}
\label{fig9}
\end{figure*}

By performing photo-luminescence excitation (PLE) spectroscopy, we identify the individual crystal-field splitting transitions within the off-resonance PL from the generated Er-O defect and we investigate the inhomogeneous linewidths and lifetimes properties of these transitions at RT.
As shown within both PLE-spectras in Figure~\ref{fig9}(a), we can identify two major wavelength-regions where distinct Er-O transitions occur from higher lying ${}^\text{4}\text{I}_{\text{13/2}}$-fields to the lower lying ${}^\text{4}\text{I}_{\text{15/2}}$-crystal fields, which also align with the observed broad peaks within the off-resonance PL-investigation, as shown in Figure~\ref{fig4}(d), (e): (1) 1534.14 nm until 1541.13 nm and (2) 1548.6 nm until 1551.66 nm. 
We observe inhomogeneous linewidths with individual full width half maximum (FWHM) values of up to 69.5 GHz $\pm$ 13.54 GHz with the narrowest FWHM identified at 11.54 GHz $\pm$ 2.58 GHz 
and 10.86 GHz $\pm$ 2.32 GHz, as illustrated in Figure~\ref{fig9}(b). This is significantly broader than previous reports, where inhomogeneous linewidths are 1 GHz \cite{weiss2021erbium} or  sub-GHz at cryogenic temperature even in SOI \cite{berkman2025long, gritsch2022narrow}. Here, the observed broadening could be attributed to the presence of oxygen, the RT investigation, the not optimal annealing and the quality of the Si material.  

The observation of wavelength-region (2), which is usually only weakly radiative when investigated at cryogenic temperatures, provides new insights into the defect at RT where FWHMs of 11.54 GHz $\pm$ 2.58 GHz and 11.03 GHz $\pm$ 3.08 GHz are determined for defects embedded into the metasurface or within unfabricated sections, respectively.
Furthermore, we observe an individual maximum resonance enhancement factor of 2.9 for a resonance at 1534.79 nm. Interestingly, by contrary we do not observe significant linewidth modification or photon-emission enhancement from resonances situated at 1548.94 nm, 1549.2 nm, 1549.48 nm and 1550.02 nm, which could point to a specific Er-site which is not interacting with metasurface.
PLE lines' lifetimes, as shown Figure~\ref{fig9}(c), of up to 2.41 ms $\pm$ 0.57 ms are observable for defects within unfabricated sections while the same resonance observed within the metasurface provides a reduced lifetime of 1.19 ms $\pm$ 0.21 ms. Not enhanced resonances in defined section (2) exhibit similar lifetimes of 1.2 ms $\pm$ 0.22 ms and 1.42 ms $\pm$ 0.39 ms for both scenarios. 

~

\section{Conclusions}
To summarize, we demonstrated collection efficiency enhancement of Er$^{3+}$ emitters at 1535 nm using SOI nanodisks metasurface. 
The metasurfaces were fabricated in SOI thin film to demonstrate a CMOS compatible monolithic integrated Erbium emission enhancement material platform. We therefore studied the PL enhancement at RT and cryogenic temperatures, showing a maximum enhancement of a factor of 5 at RT, which is consistent with an ensemble of Erbium emitters with low polarization absorption dipoles and a low previous experimental estimate of the internal quantum efficiency of approximately 2.5\% \cite{PrioloExcitation1998}. Using PLE (on-resonance excitation) we obverse longer lifetimes and a lifetime reduction in the metasurface of a factor of 2 for the 1534.79 nm transition, favored by isolating Erbium emitters radiative decay from non-radiative decay due to the metastable state during off-resonance excitation, while the FWHM of the transition is not reduced by the metasurface due to generally too low quality factors.
Our results demonstrate that SOI metasurfaces incorporating Erbium emitters can be used as monolithic integration surfaces to control the spectral and directional properties of the emitters, tailored by the metasurface's design. Our presented results open interesting opportunities for applications in integrated CMOS compatible light sources technologies in the telecom C-band at RT (e.g. luminescence thermometery \cite{sandholzer2025luminescence}), and further enhance the collection efficiency of photons from Erbium-doped nanodiodes \cite{ma16062344}, and could be applied to similar photonic platforms such as silicon carbide on insulator\cite{Baderphotoluminescence2025}. Future investigations should be directed towards improving the Mie-scattering metasurface design to achieve Purcell's enhancement, which includes considering reduced periods, various radii, with varying asymmetry providing higher Q-factors, as achieved for example in non-monolithic quasi-BIC mode slot single nanopillar\cite{KalinicQuasiBIC2023}, and understanding of the effect of closely spaced nanodisks array in enhancing the collection efficiency.
\vspace{6pt} 
\section*{Supporting Information}
This material is available free of charge online
\begin{itemize} 
\item Experimental setups; 
\item Further emitters-properties insights;
\item Measured reflection spectrum of the metasurface.

\end{itemize}

\section*{Acknowledgments}
The optical confocal characterizations have been conducted in the RMIT laboratories, partially funded by the ARC Centre of Excellence for Nanoscale BioPhotonics (No. CE140100003), and the LIEF scheme grant (No. LE140100131). 
The authors acknowledge the facilities and the scientific and technical assistance provided by RMIT University Microscopy and Microanalysis Facility, a linked laboratory of Microscopy Australia enabled by NCRIS, as well as the Micro Nano Research Facility (MNRF). This work was performed in part at the Melbourne Centre for Nanofabrication (MCN) in the Victorian Node of the Australian National Fabrication Facility (ANFF). This work was supported by RMIT University School of Engineering Support Crazy Idea EE (ID: PROJECT PLAN TASK-3-72725). S.-I.S. acknowledges the financial support provided by the JST FOREST Program (Grant No. JPMJFR203G).
S.C. acknowledges Faraz Inam and Mohammed Ashahar Ahamad for the discussion on the properties of Mie scattering broken symmetry metasurfaces. 
\section*{Authors contribution}
S.C.: Experimental design and methodology. H.A.: Metasurface Fabrication. S.-I.S.: Ion implantation. J.B., V.C.: Experimental Data acquisition, their visualization and interpretation.  J.B., S.C., H.A., V.C.: Curation, visualization, and investigation. S.C.: Supervision. J.B., H.A., S.C.: Writing original manuscript draft, reviewing and editing. All authors contributed to writing and commenting the final manuscript.

\section*{Competing Interests}
The authors have no competing interests

\section*{Data availability}
The presented data is available upon request from the corresponding author.

\bibliography{bibliography}

\end{document}